\documentclass{aa}
\usepackage[varg]{txfonts}
\usepackage{graphicx}
\pdfoutput=1
\usepackage{sistyle}
\SIthousandsep{,}
\bibpunct{(}{)}{;}{a}{}{,} 
\begin{document}
\title{Spectral variability of photospheric radiation due to faculae I:}
\subtitle{The Sun and Sun-like stars.}
\author{Charlotte M. Norris\inst{1}, Benjamin Beeck\inst{2}, Yvonne C. Unruh\inst{1}, Sami K. Solanki\inst{2,3}, Natalie A. Krivova\inst{2} \and Kok Leng Yeo\inst{2}}
\authorrunning{C.M. Norris et al.}
\titlerunning{Spectral variability of stellar photospheres due to faculae I.}
\institute{Department of Physics, Imperial College London, London SW7 2AZ, UK
  \and Max Planck Institute for Solar System Research, Justus-von-Liebig-Weg 3, 37077 G\"ottingen, Germany
  \and School of Space Research, Kyung Hee University, Yougin, 446-701, Gyeonggi, Republic of Korea }
\date{Received - / Accepted -}
\abstract {Stellar spectral variability on timescales of a day and longer, arising from magnetic surface features such as dark spots and bright faculae, is an important noise source when characterising extra-solar planets. 
Current 1D models of faculae do not capture the geometric properties and fail to reproduce observed solar facular contrasts. Magnetoconvection simulations provide facular contrasts accounting for geometry.}{We calculate facular contrast spectra from magnetoconvection models of the solar photosphere with a view to improve (a) future parameter determinations for planets with early G type host stars and (b) reconstructions of solar spectral variability.} {Regions of a solar twin (G2, $\log g=4.44$) atmosphere with a range of initial average vertical magnetic fields (100 to 500~G) were simulated using a 3D radiation-magnetohydrodynamics code, MURaM, and synthetic intensity spectra were calculated from the ultraviolet (149.5~nm) to the far infrared (160000~nm) with the ATLAS9 radiative transfer code. Nine viewing angles were investigated to account for facular positions across most of the stellar disc.} {Contrasts of the radiation from simulation boxes with different levels of magnetic flux relative to an atmosphere with no magnetic field are a complicated function of position, wavelength and magnetic field strength that is not reproduced by 1D facular models. Generally, contrasts increase towards the limb, but at UV wavelengths a saturation and decrease are observed close to the limb. Contrasts also increase strongly from the visible to the UV; there is a rich spectral dependence, with marked peaks in molecular bands and strong spectral lines. At disc centre, a complex relationship with magnetic field was found and areas of strong magnetic field can appear either dark or bright, depending on wavelength. 
Spectra calculated for a wide variety of magnetic fluxes will also serve to improve total and spectral solar irradiance reconstructions.} {}
\maketitle

\section{Introduction}
The Sun and other Sun-like stars have been found to be variable on many timescales, from minutes to years \citep{Willson1981,Radick1998,Lockwood2007,Basri2010,Basri2013,McQuillan+2012}. On the Sun, depending on the timescale, this variability is attributed to convection (granulation), oscillations (p-modes) and magnetic activity. The first two phenomena influence the variability mainly on timescales less than a day, while the magnetic field contributes predominantly on longer timescales \citep{Seleznyov2011}. Concentrations of magnetic field, seen on the surface as spots, faculae and network, vary in time as they evolve due to the constant interplay between magnetic flux and convective motions in the atmosphere. While the largest concentrations of magnetic field are spots, smaller-scale (up to 300-400~km diameter; \citet{Spruit1981,Grossman1994}) flux tubes are known as faculae. Faculae are generally bright relative to the quiet solar surface towards the limb, but have low contrasts in most wavelengths at disc centre \citep{Spruit1976}.\par 
Viewing the Sun from Earth, the radiation from different surface features is modulated due to the changing angle of observation as they move across the solar disc as the Sun rotates around its axis. Emerging and decaying features introduce additional changes. Indeed, magnetic  features are thought to be the dominating cause of solar irradiance variability on timescales of days and longer \citep{Domingo2009,Solanki2013}. 
Total solar irradiance (TSI), which is the spectrally integrated solar radiation at 1 AU, has been measured by numerous space-borne instruments since 1978 \citep{Kopp2016}. Over a solar cycle, an increase in the total solar irradiance is observed with increased magnetic activity, suggesting that bright features, i.e. faculae, dominate over dark spots on timescales of months and longer \citep{Foukal1988}. Spots, by contrast, dominate solar variability on rotational timescales. Variability on timescales of days and longer can therefore be accounted for by considering quiet sun, spot umbra and penumbra and faculae \citep{Krivova2003}.\par
Total (and spectral) solar irradiance (TSI and SSI respectively) are both important inputs into the climate system of the Earth  \citep{Haigh2007,Gray2010,Solanki2013b}. To include the effects of these magnetic features in reconstructions of the total and spectral solar variability, their contrasts with respect to the quiet photosphere must be known. Measuring facular contrasts is a difficult task as individual facular elements are generally not resolved. Most measurements are restricted to single wavelengths \citep{Auffret1990,Frazier1971,Taylor1998}. Only a comparatively small number of measurements, usually restricted to one or a few wavelengths, include information on either facular size or magnetic field \citep{Topka+1992,Topka+1997,Ortiz2002,Kobel+2011,Kobel+2012,Yeo2013}. Therefore, to account for these faculae in variability studies, that typically cover a broad wavelength range, their contrasts must be modelled to provide full spectral coverage, see e.g. \citet{Unruh1999}. \par
Facular contrasts are typically derived from semi-empirical, plane-parallel, 1D atmosphere models such as those developed by \citet{Vernazza1981,Lemaire1981,Fontenla1993,Fontenla1999,Fontenla2002,Fontenla2006}. These models adjust optical depth artificially in order to account for the longer path length traversed when viewing faculae at the limb. While these capture the overall disc-integrated properties of faculae reasonably well, they are unable to reproduce observed limb-dependent contrasts \citep{Ortiz2002,Yeo2013} as they do not take into account the geometric effects of neighbouring granules and the evacuation of the magnetic feature. This leads to a continued steep increase in the derived contrast close to the limb, which is not observed (see, e.g., \citealt{Unruh1999}). \par 
Three-dimensional simulations can provide more realistic centre to limb variations (CLVs) for facular contrasts by introducing the mix of atmospheric features present in the line of sight when viewing faculae towards the limb. Up until now, CLVs of contrasts for these features on the Sun have been modelled in 3D only for single wavelength bands, (see, e.g.,  \citealt{Afram2011}). In addition to improving CLVs of facular contrasts, this type of modelling can also provide a direct link to the magnetic flux in the facular region and thus potentially remove the need for a free parameter in solar variability models \citep{Yeo+inprep}. \par
Other Sun-like stars have been observed to vary on similar timescales as the Sun. With the advent of high precision photometric measurements, activity cycles have been observed by ground-based telescopes \citep[see, e.g.,][]{Lockwood2007,Hall2009}, and space-based telescopes such as Kepler \citep{Borucki2004} and CoRoT \citep{Baglin2003} have vastly increased the number of stars for which rotational variations have been measured  \citep{Affer2009,Basri2010,McQuillan+2014}. It is clear that solar-like rotational variability is common. These variations are expected to be caused by similar features as those seen on the Sun. \par
In exoplanet transits, if occulted by the planet, magnetic bright and dark regions appear as dips and bumps in the transit light curve respectively. If these regions are not occulted they can change the out-of-transit light level. Both these instances can cause the planet-to-star radius ratio that is obtained to differ from the real value \citep{Czesla2009}. If not taken into account, these magnetic bright and dark regions can also affect the conclusions we draw about a transiting exoplanet's atmosphere \citep{Oshagh2014,Kirk2016,Rackham2017}. Therefore, in order to include their effects in analysis, these regions must be understood for all spectral types covered by planetary transit surveys. To extend our understanding to other stars, where we are unable to resolve the surface magnetic features, and to calculate the contrasts of magnetic features, we have to rely on models or simulations. \par
In this paper, we present, for the first time, synthetic spectra from the ultraviolet (UV) to the far infrared (FIR), produced from 3D magneto-convection simulations of a Sun-like stellar photosphere with spectral type G2 and gravitational acceleration $\log g\approx4.44$. In section \ref{methods} we discuss the methods used for simulating the stellar atmosphere and producing the spectra; we also compare our results to observations of limb darkening on the Sun. In section \ref{results} we present the calculated spectra for the series of convective snapshots and detail the effects that magnetic field has on their properties. Finally, we discuss the averaged properties of the snapshots, with emphasis on the contrasts between the simulation boxes.
\section{Synthesising Spectra} \label{methods}
\subsection{MURaM simulations}
\label{methods:muram}
Solar spectral type (G2, T$_{\rm eff}\approx 5800$~K, $\log g = 4.44$) atmospheres were simulated using the magnetohydrodynamics code, MURaM \citep{Vogler2005}. The metallicity of the simulations was adopted from the ATLAS9 opacity distribution functions (ODFs) that correspond to solar values \citep{Castelli&Kurucz2001}. The simulations produce grids of temperature, pressure, density, magnetic field and velocity for 512 by 512 pixels, with 300 pixels in depth, at a horizontal and vertical grid size of 17.58~km and 10.03~km, respectively. This represents a box in the solar atmosphere of 9~Mm by 9~Mm for a depth of 3~Mm containing the photosphere and the upper layers of the convection zone. The box extends from logarithmic optical depths (at $500$~nm) of approximately $7$ to $-6$. The box has periodic boundaries in the horizontal direction and is closed at the top. The bottom boundary is open to flows of a constant entropy density. This allows for control of the effective temperature of the simulation boxes \citep{Beeckthesis}. The simulations use the OPAL equation-of-state \citep{Rogers1994,Rogers1996}. Opacity is accounted for using four wavelength bins derived from the ATLAS9 opacity distribution functions, ODFs \citep{Kurucz1993}. More detail can be found in \citet{Beeckthesis}. \par 
\begin{figure}[]
	\resizebox{\hsize}{!}{\includegraphics[]{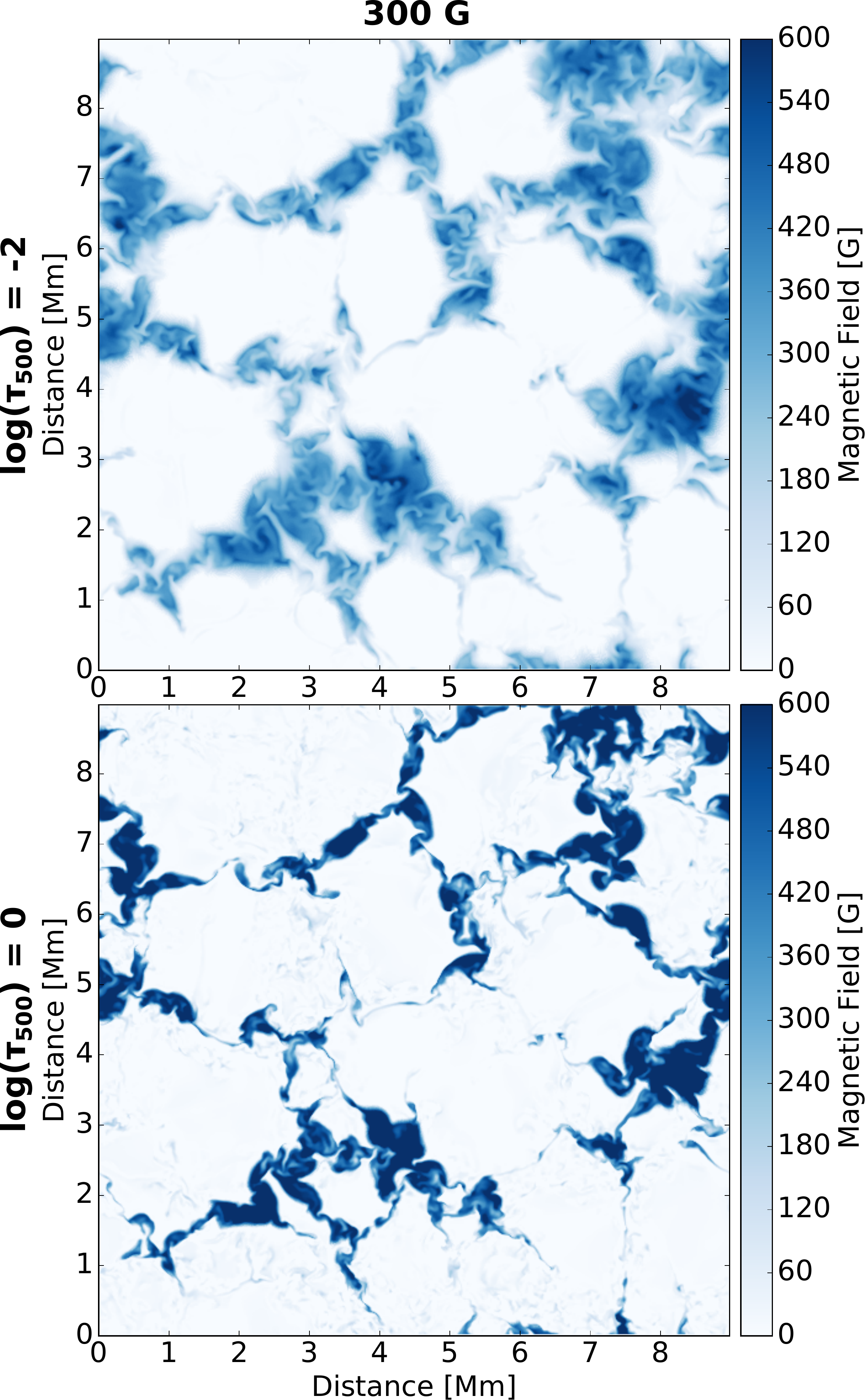}}
	\caption{The distribution of the unsigned vertical magnetic field component, $B_z$, at disc centre ($\mu=1$) for a~$\langle B_z\rangle = 300$~G simulation. The field is shown for a surface corresponding to a constant logarithmic optical depth at 500~nm, $\log(\tau_{500})$, are $-2$ (top) and $0$ (bottom).}
	\label{fig:magfield}
\end{figure}
Magnetic field-free (henceforth hydrodynamic or field-free) simulations are run for 5 hours of solar time and allowed to relax for the first 3 hours. After this relaxation time, we select ten snapshots at approximately 10 minute intervals to allow the granulation patterns to evolve sufficiently between snapshots to make them independent. The average total radiative flux of these ten snapshots is close to the average over the whole relaxed portion of the simulation. Translated into effective temperature, the standard deviation for the ten snapshots is 6~K and very similar to the value of 7~K for the whole duration of the relaxed phase \citep{Beeck2013}. Homogeneous vertical magnetic fields of 100~G, 200~G, 300~G, 400~G and 500~G (chosen in order to represent a range of activity levels) are then injected into one of the hydrodynamic snapshots. On the Sun, areas of this size with an average magnetic field of 500~G are not so common. They are mainly included to allow for comparison with stars of different activity levels. For each initial average vertical magnetic field, $\langle B_z\rangle$,  the simulations are allowed to relax for two hours of simulated stellar time. Snapshots are taken at approximately 7-minute intervals to allow for sufficient evolution of the granulation pattern. As in the case of the field-free simulations, the average total radiative flux for a given magnetisation is close to the average over the whole relaxed portion of the simulation. We limit the number of snapshots to 10 for each of our chosen average magnetic fields as the number of surface elements is large and synthesising spectra is time costly.\par
Figure~\ref{fig:magfield} shows the vertical magnetic field, $B_z$, in a snapshot of a simulation that had an initial average vertical magnetic field of 300~G. The magnetic field is shown for constant optical depths at 500~nm, $\tau_{500}$, of 1 (bottom panel), and 10$^{-2}$ (top panel). The field is concentrated in the intergranular lanes due to the convective motions of the plasma. The vertical magnetic field strength within the granules is low, e.g. less than 15~G at $\tau_{500}=1$ in the $\langle B_z\rangle=300~G$ box. The field is more diffuse higher in the atmosphere as the surrounding gas pressure in the atmosphere is lower. \par

\begin{figure*}[]
	\centering
	\resizebox{\hsize}{!}{\includegraphics[]{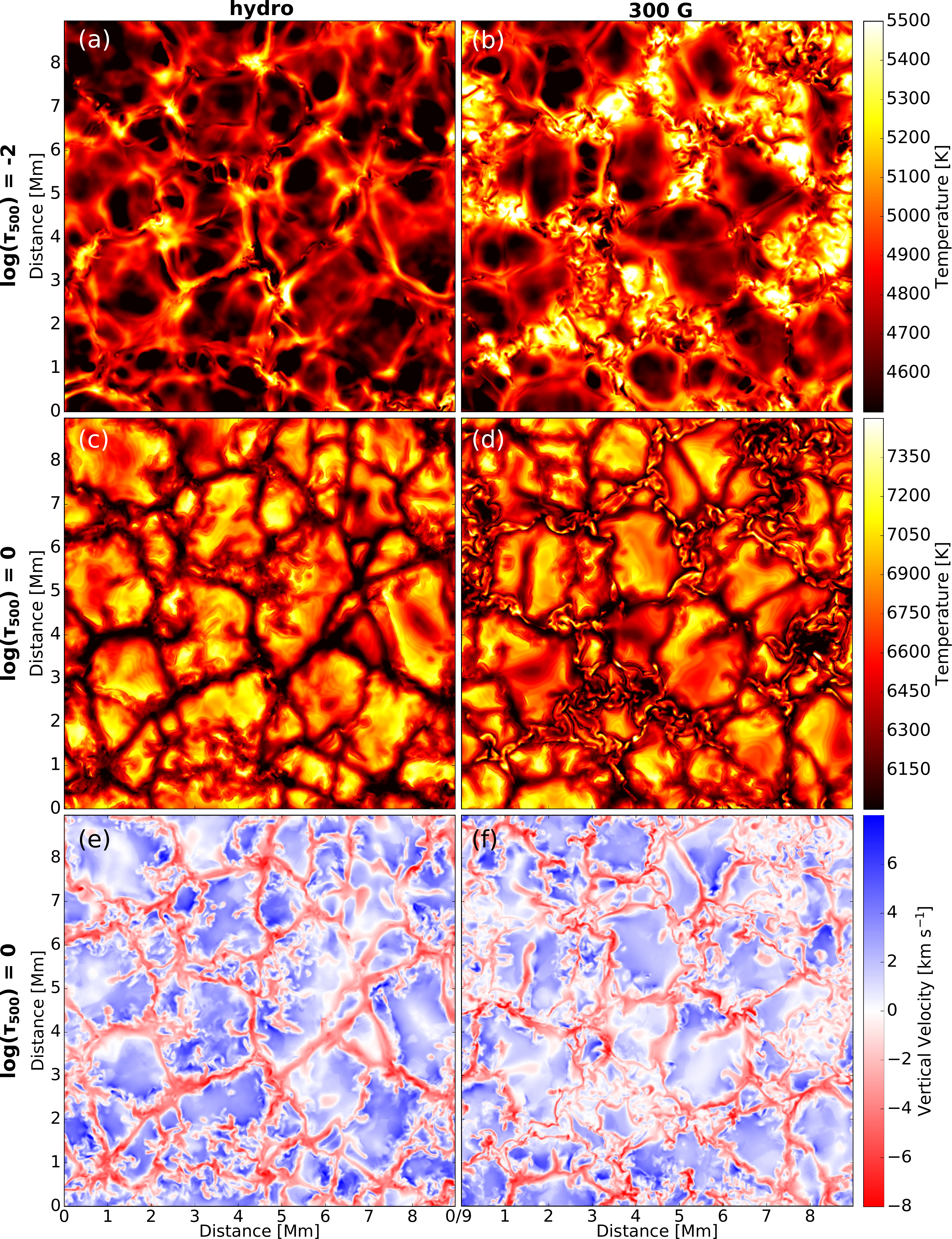}}
	\caption{The temperatures (a-d) and vertical velocities (e and f) at disc centre for a field-free simulation (left) and a~~$\langle B_z\rangle=300$~G simulation (right). The temperatures and velocities are shown at a constant logarithmic optical depth at 500~nm, $\log(\tau_{500})=-2$ (top) and $\log(\tau_{500})=0$ (middle and bottom).}
	\label{fig:temp}
\end{figure*}
To illustrate the effect of the magnetic field on the convection, Figure~\ref{fig:temp} shows the temperatures for a hydrodynamic box (panels a and c) and a 300~G box (panels b and d). The panels show the same constant optical depth selection as in Fig.~\ref{fig:magfield}: $\log(\tau_{500})=-2$ (a and b) and $\log(\tau_{500})=0$ (c and d). The snapshot displayed in the ${\langle}B_z\rangle=300$~G image is the same as shown in Fig.~\ref{fig:magfield}. Images of the vertical velocities at a constant $\tau_{500}$ of 1 are displayed in Fig.~\ref{fig:temp} e and f in order to show the positions of the up-welling convective granules and the down-flowing inter-granular lanes. A granular structure can be seen in both hydrodynamic and ${\langle}B_z\rangle=300$~G images. At $\log(\tau_{500})=0$ the granules are hot and the temperature is cooler in the intergranular lanes. In the magnetic run, a larger number of high-temperature features are observed compared to the field-free simulations. These additional temperature peaks tend to align with peaks of the  magnetic field in the intergranular lanes. However, there are some exceptions where large areas of strong magnetic field appear cooler than granules \citep{Spruit1976}. In the upper layers of the atmosphere we begin to see reversed granulation, i.e. at $\log(\tau_{500})=-2$ the centre of granules appear cooler than their edges.The higher atmospheric layers for both the no-field and the 300~G simulation have a more diffuse distribution than the deeper layers, as seen for the magnetic field. Detailed investigations into the structure of MURaM simulations and how they change when a magnetic field has been introduced are shown by \citet{Beeck2013,Beeck2015}.

\subsection{Spectral Synthesis} \label{SpectralSynthesis}
We calculate emergent intensity spectra along rays that are laid through each pixel of the simulation box. To model different viewing angles, corresponding to different positions on the star, these rays were inclined at angles $\theta$ to the local normal ($\theta$ is the heliocentric angle). Specifically, we compute rays with inclinations corresponding to limb distances of $\mu=\cos\theta$ between $0.2$ and $1.0$ in steps of $0.1$. For ease of visual interpretation, feature positions were maintained in images of emergent intensity by pivoting the sight lines at the mean geometrical depth, $\langle z \rangle$, where $\tau_{500}$, the disc centre optical depth at $500$~nm, is unity. \par
MURaM provides temperature, pressure, density, magnetic field, velocity and geometric depth for each each grid point. The atmospheric quantities at the locations of the rays are obtained by performing a 2D linear interpolation along the line of sight for each pixel with a vertical resolution equal to that of the simulations (10.03~km). Double resolution was tested and found to have negligible effect on the calculated emergent intensities. Along each ray, the column mass, electron number density and continuum absorption coefficient at 500~nm were derived for use in the ATLAS9 spectral calculation. The atmospheric parameters along the sight lines were then linearly interpolated onto a 256 point grid. A range of upper limits and distributions of optical depths were tested for calculating emergent intensities. An evenly spaced grid in logarithmic optical depth with values between $\log(\tau_{500})=-5.0$ and $2.5$ was selected as this is where most of the radiation is produced. Also, this range of optical depths is completely contained within the simulation box at each spatial pixel. \par
To calculate the emergent intensities, the radiative transfer code ATLAS9 \citep{Kurucz1992,Castelli1994} was used. ATLAS9 allows for fast and accurate computation of continuum intensities across a wide wavelength range. Spectra were synthesised for each sight line of the MURaM box. This calculation is done under the assumption of local thermodynamic equilibrium (LTE) using an opacity distribution function (ODF) table \citep{Castelli&Kurucz2001} to account for the opacity of the millions of atomic and molecular spectral lines in a stellar atmosphere, particularly in the UV. The ODF table used was calculated for solar metallicity with a microturbulence of 1~km/s. This is the nearest tabulation to the RMS speeds found for the simulations at an optical depth of unity. For each pixel, we calculate spectra ranging from 149.5~nm to 160~$\mu$m. The spectra are calculated for 1040 wavelength bins that correspond to the resolution provided by the ATLAS9 ODFs. The resolution of the spectra thus ranges from 1~nm in the UV, to 2~nm in the visible, and up to 20~$\mu$m in the far infrared. \par 
ATLAS9 assumes LTE for its calculations. For wavelengths below 300~nm the absolute values of emergent intensity begin to deviate significantly from the values found when allowing for non-LTE. However, relative values continue to be reasonably accurate down to 164~nm \citep{Krivova2006}. Thus, for the purpose of this paper, only wavelengths above 164~nm are discussed.\par
\begin{figure*}[]
	\centering
	\resizebox{\hsize}{!}{\includegraphics[]{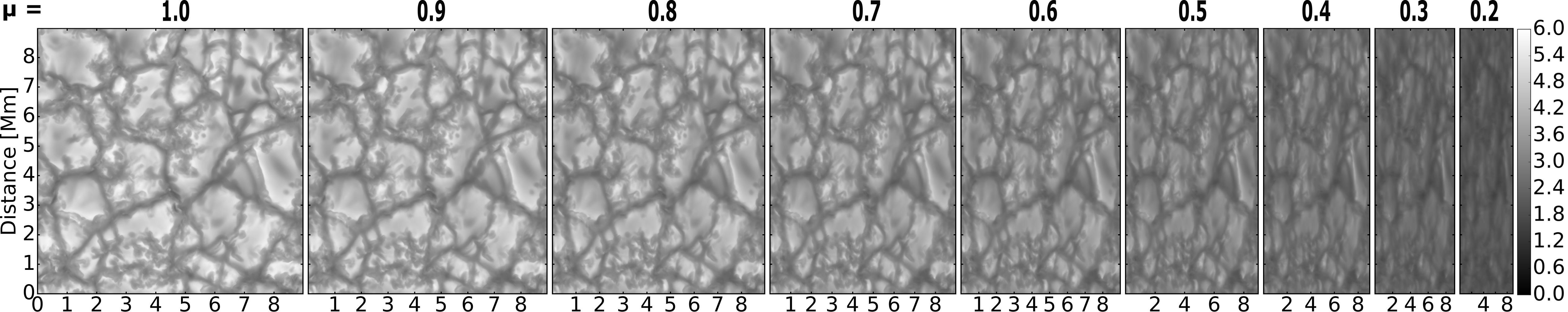}}
	\caption{Emergent intensities (10$^{-8}$~W~m$^{-2}$~sr$^{-1}$~Hz$^{-1}$) at a wavelength of 601~nm for a simulated hydrodynamic snapshot viewed at limb distances ranging from $\mu$~=~1.0 in the left most image, to $\mu$~=~0.2 in the right most image. The distance shown is the distance on the surface of the Sun in Mm, the boxes are foreshortened in the horizontal direction to show the size of the box observed on the solar disc. The employed grey-scale is indicated on the right of the figure.}
	\label{fig:LDimages}
\end{figure*}
\begin{figure}[]
\resizebox{\hsize}{!}{\includegraphics[width=\textwidth]{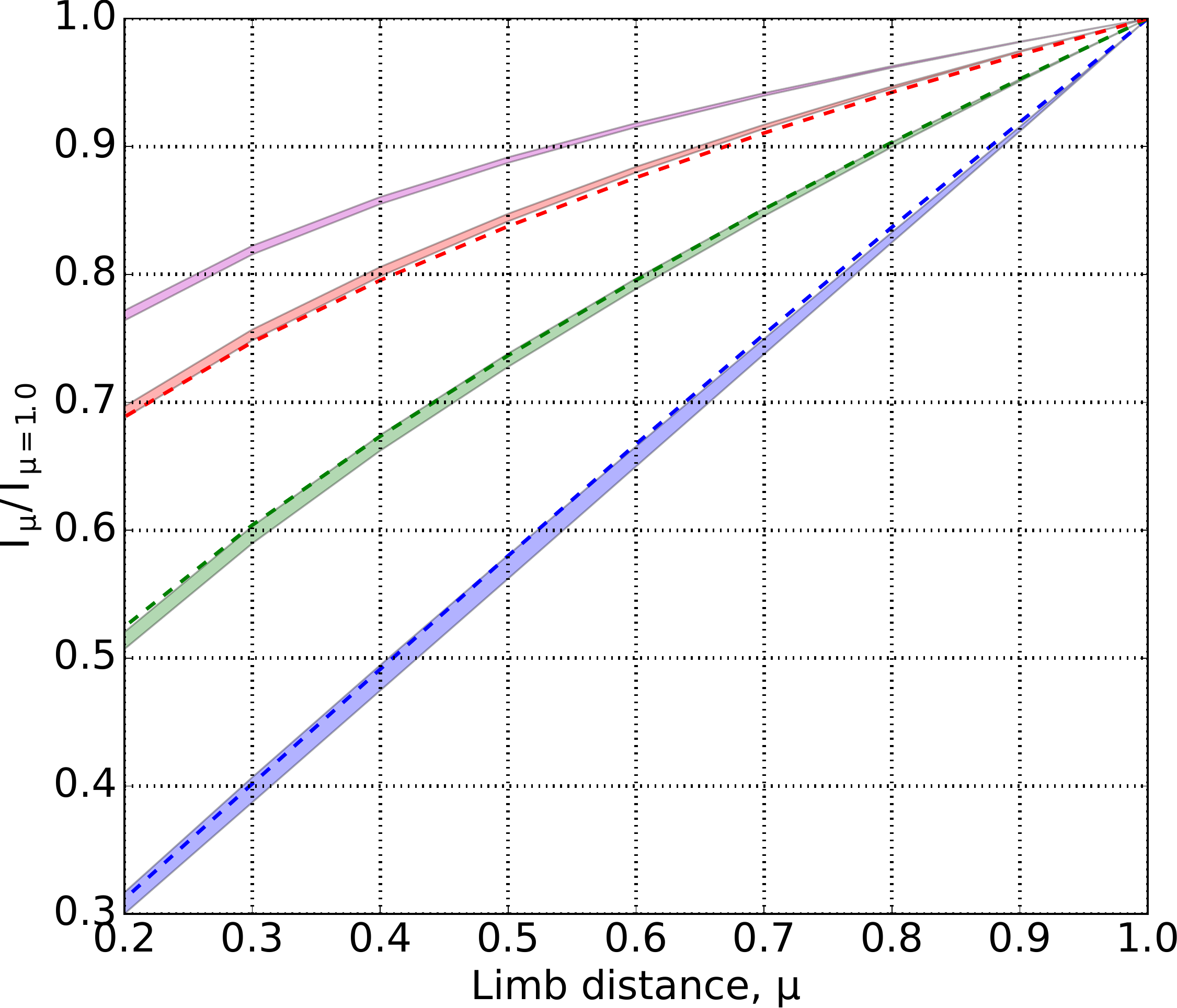}}

\caption{Comparisons of the observed limb darkening polynomials \citep{Neckel1994} (shown by dashed lines at wavelengths of 390.9~nm (blue), 611.0~nm (green) and 1099.0~nm (red)) with MURaM-derived limb darkening. The shaded regions show the average intensity of snapshots as a function of limb distance, $\mu$, between the hydrodynamic values (lower limit) and the 100~G snapshots (upper limit) for various wavelength bins centred on 391.0~nm (blue), 611.0~nm (green), 1097.5~nm (red) and 1597.5~nm (magenta). The intensities have been normalised to unity at disc centre. There are no published measurements of limb darkening above 1098~nm.}
	\label{fig:G2_hydro_totalmean_N&Lcomp}
\end{figure}
Figure~\ref{fig:LDimages} shows the emergent intensities at a wavelength of $601$~nm for a single hydrodynamic snapshot seen at a range of viewing angles from disc centre ($\mu=1.0$) to near the stellar limb ($\mu=0.2$). Darkening of the images can be seen towards the limb, as expected in the visible, and a foreshortening of the horizontal observed distance is shown to demonstrate the box's appearance on the solar disc. \par
We confirmed that the mean spectra derived from the field-free simulations reproduce observed limb darkening laws, by comparing to limb darkening polynomials previously derived from observations by \citet{Neckel1994}. These polynomials are tabulated for wavelengths between 303~nm and 1099~nm. In Fig.~\ref{fig:G2_hydro_totalmean_N&Lcomp} shaded regions indicate the limb darkening derived from non-magnetic convection simulations (lower bounds) and the 100~G snapshots (upper bounds) for wavelengths of 391~nm (blue), 611~nm (green), 1097.5~nm (red) and 1597.5~nm (magenta) and are compared to the Neckel \& Labs tabulations at 391~nm (blue), 611~nm (green) and 1099~nm (red) in dashed lines. Overall, the calculated and observed limb darkening laws agree well, although, in the visible, the limb darkening gradient of the polynomials is slightly shallower than  than that of the MURaM-derived hydrodynamic values. These differences may in part be accounted for by the fact that we never observe the truly field-free quiet sun. When a magnetic field is present, the appearance of faculae, that are bright at the limb and have low contrast at disc centre \citep{Spruit1976}, leads to atmospheres that have a shallower limb darkening gradient than those that are field-free. This is seen in the $\langle B_z\rangle=100$~G simulations shown by the upper limit of the shaded regions in Fig. \ref{fig:G2_hydro_totalmean_N&Lcomp}. At some wavelengths, particularly towards the UV, missing opacity sources may also contribute to the differences between the simulations and the measurements, although these should have a minor effect when the intensities are normalised, as they are in Fig.~\ref{fig:G2_hydro_totalmean_N&Lcomp}.
\section{Results} \label{results}
\subsection{Synthetic spectra}
\begin{figure*}[]
	\centering
\textit{}	\resizebox{\hsize}{!}{\includegraphics[width=\textwidth]{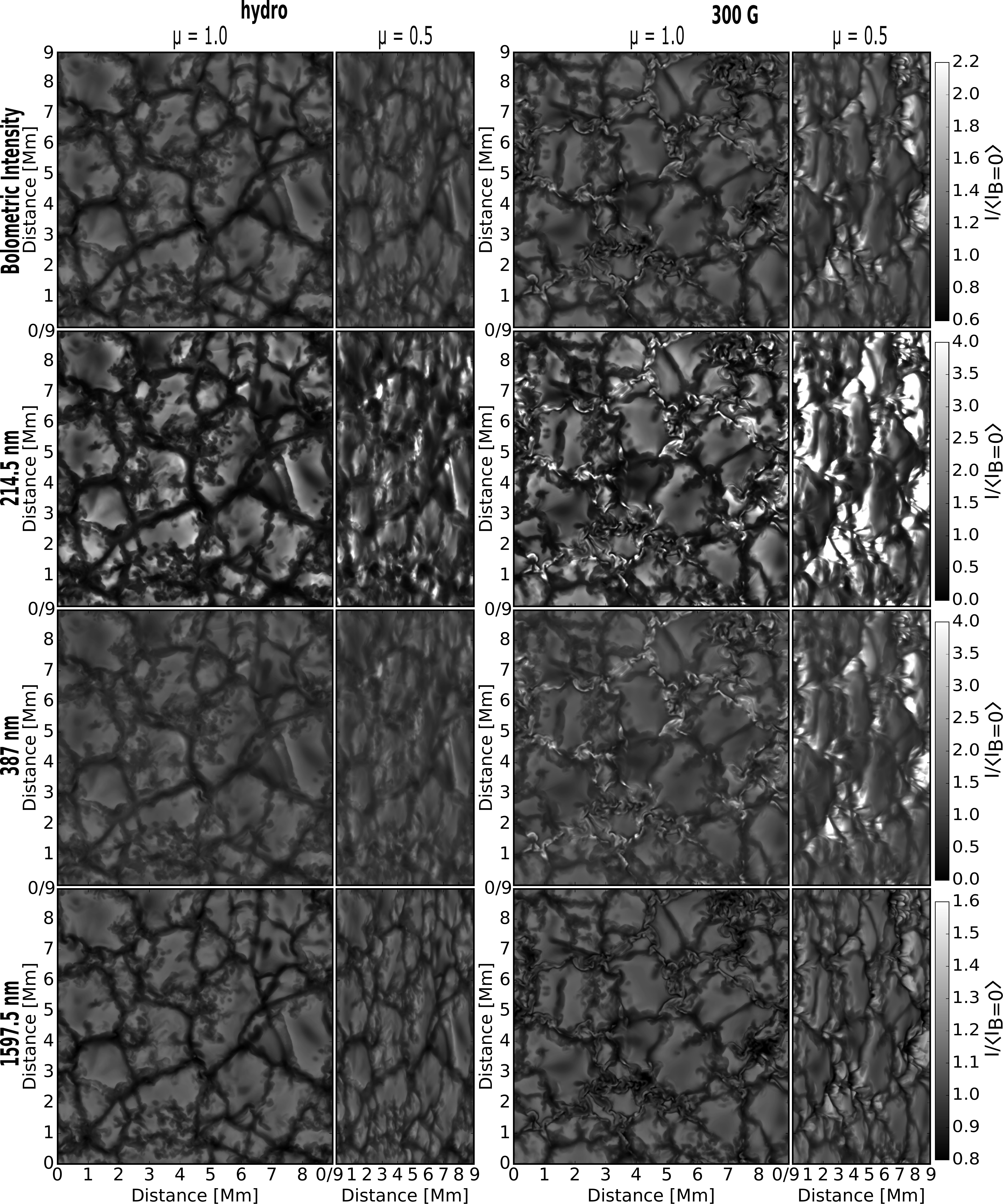}}
	\caption{Images of the emergent intensities from a G2 simulated atmospheric snapshot divided by the mean intensity of the field-free simulations for the same wavelengths and limb angles. The images are shown for a field-free snapshot (left column) and $\langle B_z\rangle=300$~G (right column). Two limb distances are shown in each column $\mu=1.0$ (left) and $0.5$ (right). Images from top to bottom: bolometric intensity and intensities for the 214.5~nm, 387~nm and 1.5975~$\mu$m ATLAS9 wavelength bins. These snapshots are the same as shown in Fig.~\ref{fig:temp}. Note the different normalised intensity ranges for the  different wavelength bins (grey-scale bars on the right). Additional figures \ref{fig:spectralimages100200} and \ref{fig:spectralimages400500} in Appendix A show equivalent images from the remaining initial magnetic field simulations.}
		\label{fig:spectralimages}
\end{figure*}

In this section we present the calculated synthetic spectra. Figure~\ref{fig:spectralimages} shows images of the emergent intensities from example snapshots. The two columns compare images of a hydrodynamic simulation (left column, the same snapshot as in Figs.~\ref{fig:temp}a, \ref{fig:temp}c, and \ref{fig:LDimages}) and a 300~G simulation (right column, the same snapshot as in Figs.~\ref{fig:magfield}, \ref{fig:temp}b and \ref{fig:temp}d). Each column contains images at two different viewing angles ($\mu$~=~1.0, left, and 0.5, right). Images are displayed for bolometric intensity (top) and for multiple ATLAS9 wavelength bins centred on 214.5 nm, 387 nm and 1597.5 nm. These wavelengths were selected to represent the range of intensities and contrasts seen across wavelengths. Faculae can appear bright or dark at different wavelengths. The bolometric intensity is plotted in order to display the overall  brightness. 
Visible wavelengths are not shown in this figure, as the irradiance in the visible dominates the bolometric intensity, and therefore would not provide any additional information. These images have been divided by the mean intensity of all pixels in the hydrodynamic simulations at each limb angle. This removes the effect of limb darkening and allows direct comparison of the differences in structure seen at different viewing angles as well as between non-magnetic and magnetic snapshots. Colour scales vary between wavelength bins so that structures can be seen more easily. \par
Faculae can be seen to appear between the convective granules in the 300~G images, whereas in the hydrodynamic images only bright granules and dark inter-granular lanes are seen. These facular regions align with the regions of concentrated magnetic field seen in Fig.~\ref{fig:magfield}. The 3D nature of the solar atmosphere, and particularly the facular regions can be clearly seen in the $\mu=0.5$ images. The shapes of the bright areas change as one views towards the limb. Parts of the faculae are obscured by the granules and the brightness of these regions increases as the hot walls of the faculae come into view \citep{Spruit1976}. The hot walls correspond to the sides of the background granules \citep{Carlsson2004,Keller2004}. For a given simulation, the features observed remain the same throughout the wavelength range, however their relative brightness clearly changes. The magnetic bright regions stand out more at shorter wavelengths (see also Fig.~\ref{fig:contrastspectra}) and in  molecular bands such as the CN violet band or the G band where the opacity decreases with increasing temperature, allowing the hot magnetic features to stand out even when seen at disc centre (see, e.g.,  \citet{Schuessler+2003,Shelyag2004,Sanchez-Almeida+2001,Shapiro+2015}. \par
It is noticeable that the brightest faculae, at both plotted $\mu$ values, have contrasts in excess of a factor of 4 of the average field-free value at $214$~nm, whereas at $387$~nm this value is reached only at $\mu=0.5$. This is in qualitative agreement with the results of the SuFI imager \citep{Gandorfer+2011} on the Sunrise balloon-borne solar observatory \citep{Solanki+2010,Barthol+2011,Berkefeld+2011}, which delivered seeing-free, high-resolution UV images with low scattered light. Thus \cite{Riethmueller+2010} found contrasts of up to a factor of 5 at 214~nm, but only up to a factor of 2.4 at 388~nm. Note that the observational values are somewhat lower than in the magnetoconvection simulations due to the influence of the point-spread function of the telescope on the observations \citep{Riethmueller+2014}.\par
In the 300~G snapshot, and in snapshots with larger initial vertical magnetic fields (shown in Appendix \ref{App:AddFigs}), additional dark magnetic features, larger than faculae, can be seen at $\mu=1$. These are small pores and have a broadly similar magnetic structure to that of facular magnetic concentrations in that both are often described by flux tubes or flux sheets \citep{Spruit1981}.\par
\begin{figure*}[]
	\resizebox{\hsize}{!}{\includegraphics[width=\textwidth]{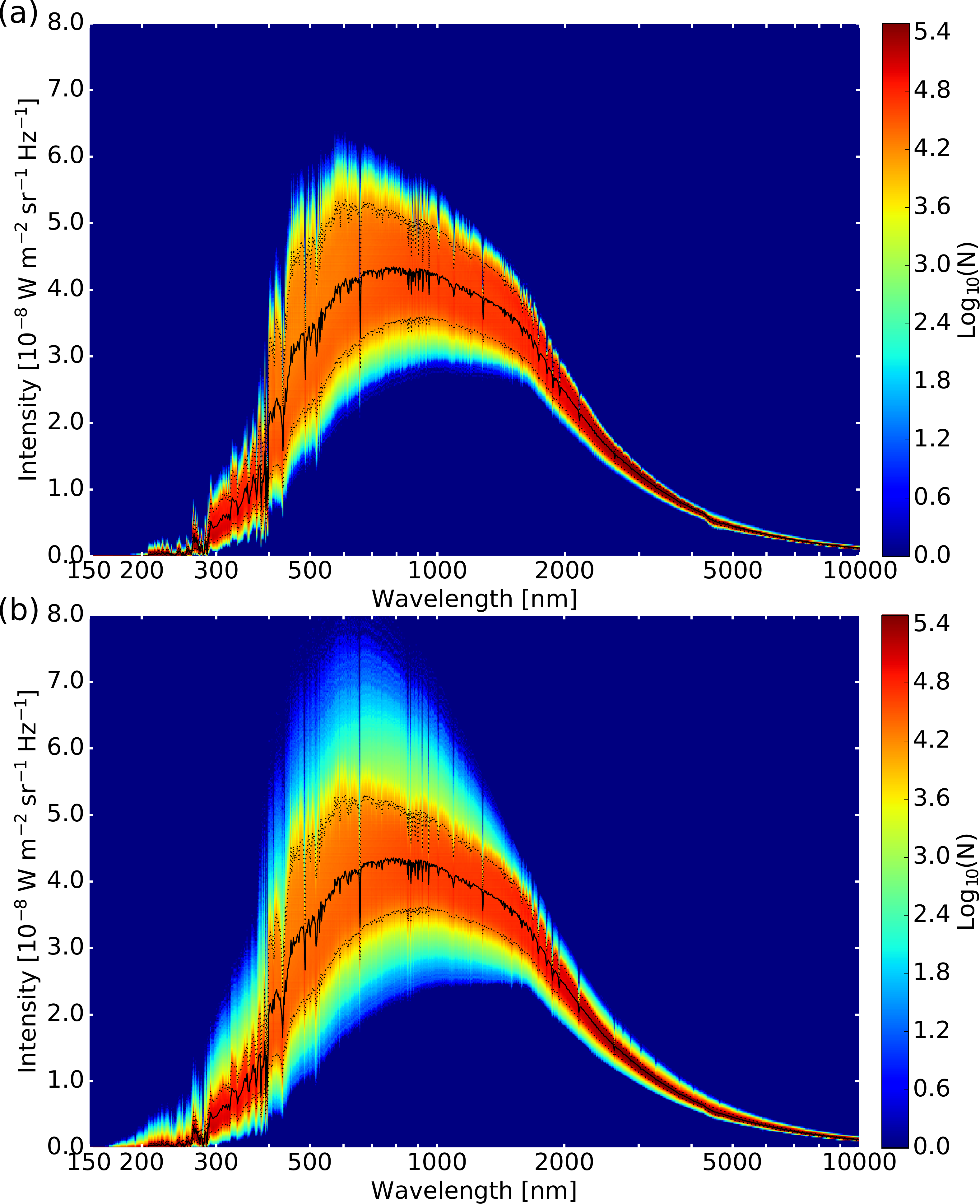}}
	\caption{Histograms of intensity in each ATLAS9 wavelength bin for all pixels in all snapshots for a given $\langle B_z\rangle$ at disc centre with the mean intensity spectra of all pixels in all corresponding simulations overlaid in a solid black line and the 5th and 95th percentile overlaid in dotted black lines. The intensity bin size is $2 \times 10^{-10}$~W~m$^{-2}$~sr$^{-1}$~Hz$^{-1}$. The wavelength bins are various sizes, as detailed in section \ref{SpectralSynthesis}, and are displayed in log space.}
	\label{fig:2dhistogram}
\end{figure*}
A histogram of the intensities between 150 and 10\,000~nm in the ATLAS9 frequency bins of all 2\,621\,440 pixels from all ten disc centre hydrodynamic snapshots is shown in the top panel of Fig.~\ref{fig:2dhistogram}. The same type of histogram for all ten disc centre snapshots with $\langle B_z \rangle=300$~G is shown in the bottom panel. Logarithmic values of the counts of pixels in each bin are indicated by the colour bar. Each histogram is overlaid with a solid black line displaying the mean intensity spectrum of all pixels in the corresponding set of simulations (same $\mu$ and $\langle B_z \rangle$); dotted lines indicate the 5th and 95th percentile of the pixel intensities. The histograms reveal the distribution of emergent intensities across a snapshot to be asymmetric and highly non-Gaussian.\par 
A broadening of the histogram in intensity can be seen in the visible wavelengths, where the mean intensity is largest. Much less spread in absolute intensity is seen at longer wavelengths, and in the UV. This is due to the lower values of emergent intensities at these wavelengths. The spread of intensities relative to the mean value is in fact larger in the UV than in the visible. This will be discussed in more detail later in this section.
The histograms tend to be narrower at positions where spectral lines appear in the mean spectrum, e.g., at H$\alpha$ ($\sim 656$~nm). These spectral lines lower the intensity significantly and are strongly saturated, i.e. even large fluctuations in temperature will not produce significant changes in the line core intensities, leading to less variation in intensities across the simulations.\par
\begin{figure*}
\sidecaption
	\includegraphics[width=12cm]{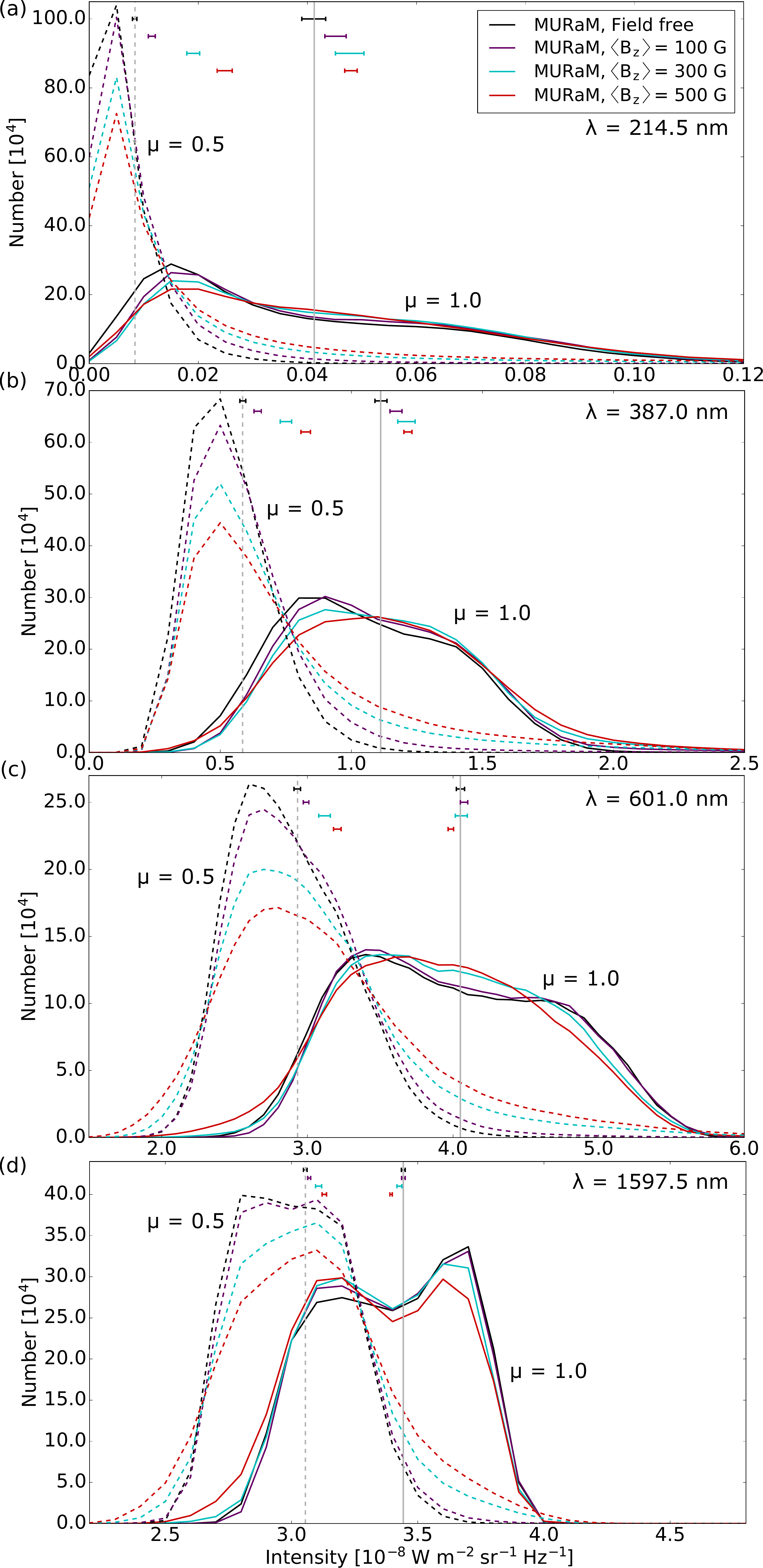}
	\caption{Histograms of intensity for all pixels in all snapshots for a given $\langle B_z\rangle$ shown in the$~$legend. The field-free mean intensities (averaged over all field-free snapshots) are marked by grey vertical lines with the minimum and maximum values of the mean intensities of the individual field-free snapshots indicated by black horizontal bars. Also shown are histograms and the range of snapshot-averaged intensities for simulations with $\langle B_z\rangle$ of 100~G (purple), 300~G (cyan) and 500~G (red). The histograms are shown at two different limb positions: $\mu=1.0$ (solid curves) and $\mu=0.5$ (dashed curves). Figure~\ref{fig:hist}a shows histograms for a wavelength of 214.5~nm with a bin width of $5\times 10^{-11}$~W~m$^{-2}$~sr$^{-1}$~Hz$^{-1}$. Figures \ref{fig:hist}b, c and d show wavelengths of 387, 601 and 1597.5~nm respectively, with a bin width of $1\times 10^{-9}$~W~m$^{-2}$~sr$^{-1}$~Hz$^{-1}$. Both, the $x$ and the $y$ scale vary between graphs.}
	\label{fig:hist}
\end{figure*}
The tails of the 300~G snapshot histograms are extended in intensity compared with the hydrodynamic simulations. Note the more extended green-blue shaded region in the lower frame of Fig. 6 compared with the upper frame. Because of the logarithmic scale, these wings correspond to relatively few pixels. This is illustrated in more detail in Fig.~\ref{fig:hist} where we show example histograms at two limb distances ($\mu=0.5$ and $1.0$) at four different wavelengths for the non-magnetic (black), 100~G (purple), 300~G (cyan) and 500~G (red) magnetic snapshots. The field-free mean intensities for each limb distance are indicated by vertical grey lines. Minimum and maximum values from the mean intensities over each of the 10 single snapshots are shown in the corresponding colour as error bars. The three panels show, from top to bottom, histograms for wavelength bins centred on 214.5~nm, 387~nm, 601.0~nm and 1597.5~nm (referred to as 1.6~$\mu$m), i.e., in the UV, visible and NIR.\par 
At disc centre, the intensity distribution of the non-magnetic snapshots clearly shows the signature of convection: the distributions for all of the wavelengths shown are double peaked with the peak at lower intensities capturing predominantly pixels in the dark intergranular lanes and the peak at higher intensity values capturing pixels in the bright granules. At 601~nm, the dark and bright peaks are centred at intensities approximately 15\% below and above the mean intensity in the non-magnetic snapshots (see solid black line in Fig.~\ref{fig:hist}(c)). As $\langle B_z \rangle$ increases, the distributions tend towards a more centrally peaked shape, with slightly broader tails at lower intensities. This is consistent with results found by \citet{Afram2011}. The field free and magnetic simulations have comparable disc centre intensities with minimal extension in the tails (compared to at the limb) because small-scale features here have low contrast relative to the quiet Sun, therefore very little dynamic range is added. \par
Similar behaviour is seen at disc centre for the CN violet band at approximately 387~nm, though the histogram distributions are wider and show more extended high-intensity tails, especially for simulations with $\langle B_z \rangle \geq 300$~G. The intensity distribution at 214.5~nm, relative to the mean intensity of the boxes, is very broad. The dominant lower-intensity peak (at approximately 60\% of the mean intensity, $\langle I_{215}\rangle$) weakens gradually with increasing magnetic field. The secondary bright peak (centred near $2\langle I_{215}\rangle$) is very extended with tails stretching to beyond $4\langle I_{215}\rangle$. This large spread is due to the high temperature sensitivity of the radiation at short wavelengths and the large number of spectral lines present in the UV.
In parts of the NIR where we can see to deeper atmospheric layers due to the H$^-$ opacity being at a minimum for these wavelengths, the disc-centre histograms are strongly double peaked with the brighter peak dominating, as illustrated by the histograms at 1597.5~nm shown in Fig.~\ref{fig:hist}(d). This difference between the histograms in the visible and NIR is likely due to the narrowing of intergranular downflow lanes with depth. \par
At smaller $\mu$ (dashed lines in Fig.~\ref{fig:hist}) the distributions for the wavelengths shown here are shifted to lower intensities due to limb darkening. The distributions become more single peaked with narrower central widths but more extended tails, particularly for the brighter intensities. Although the width of the peaks at $\mu=0.5$ are narrower than at disc centre, the largest intensities seen in the tails match, or exceed, those seen at disc centre. At 1.6~$\mu$m, the distribution retains its double-peaked shape further towards the limb and can still be seen for small $\langle B_z \rangle$ at $\mu=0.5$. This wavelength also shows much less of a high intensity tail than the other ones displayed.\par 
Towards the limb, the counts in the peaks for all wavelengths decrease significantly with increasing $\langle B_z \rangle$: the tails are lengthened both towards brighter and darker intensities, though with a larger effect on the higher intensities. This effect of lengthening tails with increasing $\langle B_z \rangle$ is due to the magnetic field introducing a larger variety of atmospheric structures. As the magnetic field increases, larger and more numerous magnetic features begin to appear. Hot walls associated with the magnetic features are more visible towards the limb of the star which gives rise to the lengthened tails at high intensities. Dark lanes arise by the faculae due to the cool gas above the adjacent granules and in the magnetic flux tube dominating the radiation in these pixels and increasing the length of the low-intensity tail as seen in \citet{Keller2004} and \citet{Carlsson2004}. \par
For a given mean magnetic field, the shape of an individual intensity histogram is very similar to that of the full histogram that combines all snapshots. The mean intensity value, however, changes slightly from one snapshot to another, reflecting the evolving granulation. The minimum and maximum values shown in Fig.~\ref{fig:hist} give the range of temporal variations of the mean intensities of the ten snapshots of a given initial $\langle B_z \rangle$. We caution that these ranges are based on subsets of ten snapshots, though for each subset, the mean effective temperature and its standard deviation are representative of the full relaxed phase of the simulations (see Sec.~\ref{methods:muram}).\par
The temporal variations of the mean intensity are much smaller than the spatial variations displayed within a snapshot. Relative to the mean intensity, the temporal variations show very little dependence on  the viewing angle or mean magnetic field.
Variability is larger for shorter wavelengths, particularly in the UV, due to the larger dependence on temperatures of the Planck function at shorter wavelengths. The many spectral lines and the higher layers of the atmosphere from which the radiation arises also contribute.

\subsection{Contrasts of magnetic snapshots}
As faculae often appear in groups and can rarely be fully resolved in observations, contrasts were calculated using all pixels of the snapshots, including those in the granules, i.e. with small magnetic field strengths. Contrasts were calculated based on the mean spectra of all pixels in all MURaM simulations of a given $\langle B_z\rangle$ and $\mu$ relative to the total mean over all the pixels in all hydrodynamic boxes at the same limb distance, $\mu$, 
\begin{equation} \label{eq:1}
C(\langle B_z\rangle, \lambda,\mu)\,=\, \frac{I(\langle B_z\rangle, \lambda, \mu) - I(0,\lambda,\mu)}{I(0,\lambda,\mu)}.
\end{equation}
$C(\langle B_z\rangle,\lambda,\mu)$ is the spectral contrast for a given average vertical magnetic field, $\langle B_z\rangle$, and limb angle, $\mu$; $I(\langle B_z\rangle, \lambda,\mu)$ is the mean spectral intensity over all pixels in all simulations for $\langle B_z\rangle$, at a given $\mu$; and $I(0,\lambda, \mu)$ is the mean spectral intensity over all pixels in all hydrodynamic boxes at a given $\mu$. This approach was chosen with stellar applications in mind, where the spatial resolution is low (at best) and hence spatially averaged intensities and contrasts are more appropriate. It is worth noting that by averaging over a large area, we take into account not only the bright component of the magnetic features, but cover all the types of magnetic features that are created for a given spatially averaged magnetic field. Therefore, in that sense our "facular" contrasts for the large flux simulations can be somewhat different from what stellar astronomers traditionally consider them to be, i.e. the purely bright component.\par
\begin{figure*}
\sidecaption
\includegraphics[width=12cm]{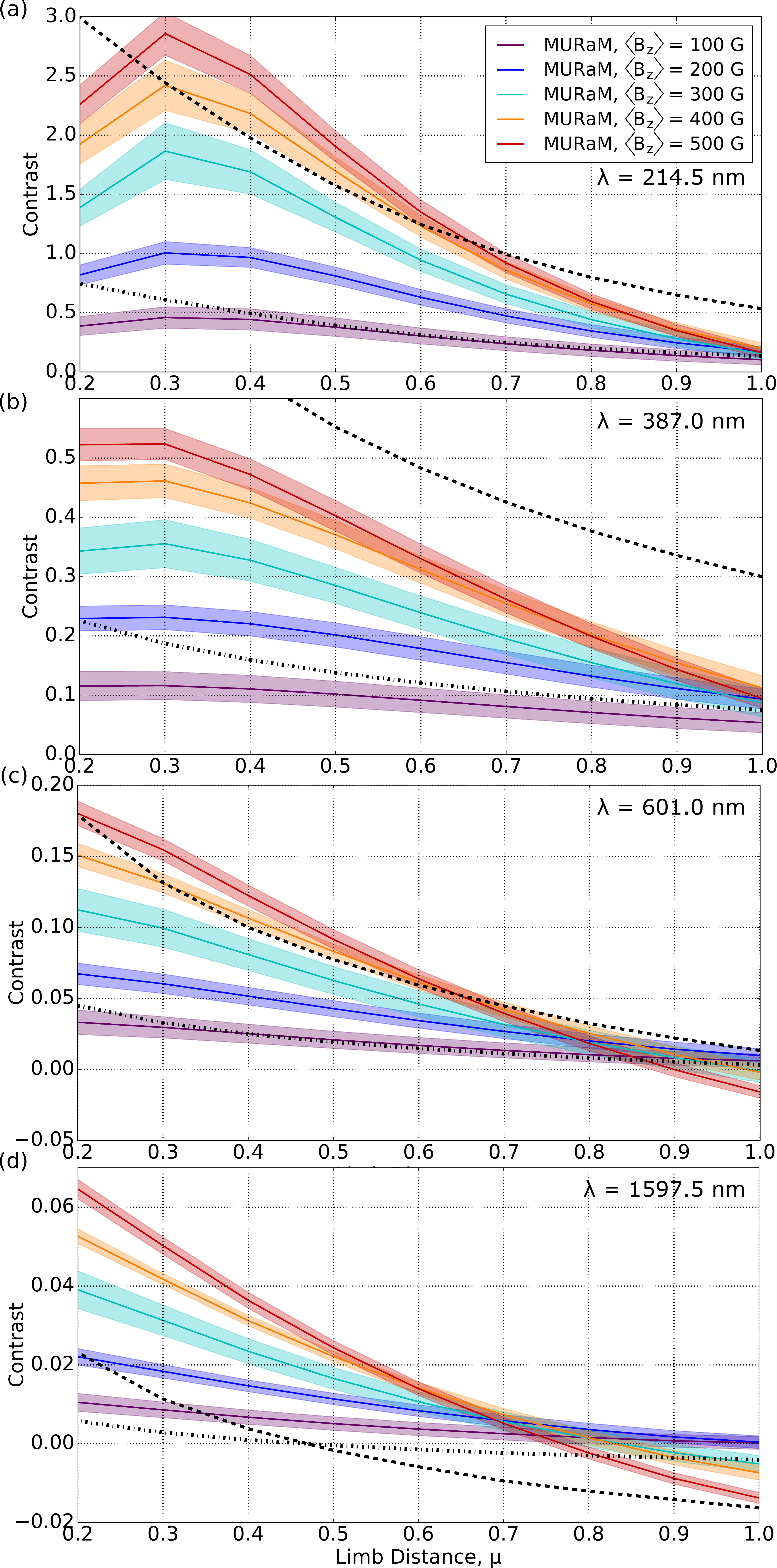}
	\caption{The average contrast vs limb distance, $\mu$, of all pixels for all snapshots of a given $\langle B_z\rangle$ relative to the average intensity of the hydrodynamic snapshots. These are shown for 214.5~nm (a), 387~nm (b), 601~nm (c) and 1.5975~$\mu$m (d). The shaded regions show the spread calculated from the standard deviations of the means of the snapshots at each $\mu$, incorporating both the magnetic and hydrodynamic spreads. Colours corresponding to $\langle B_z\rangle$ are given in the legend of the figure. The dashed and dash-dotted lines show contrasts derived for 1D facular model (see text) with a 100$~\%$ and a 25$~\%$ filling factor, respectively.}
	\label{fig:contrastvslimb}
\end{figure*}

\begin{figure*}
	\centering
    \includegraphics[width=\textwidth]{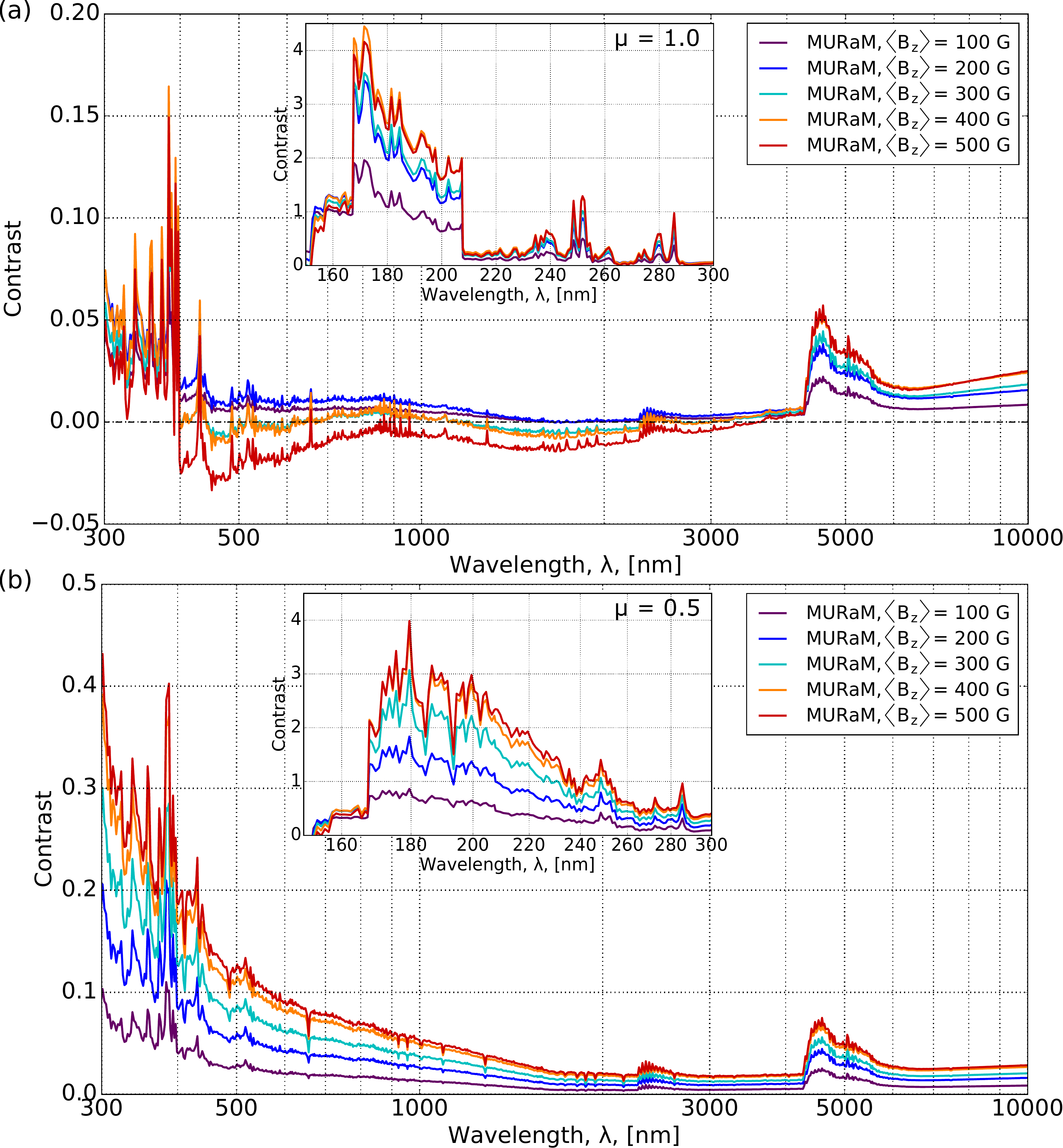}	
    \caption{The average contrast spectra of all pixels for all snapshots of a given $\langle B_z\rangle$ relative to the average of the hydrodynamic snapshots shown for a wavelength range of 150~nm to 10\,000~nm. The main plots show wavelengths of 300~nm to 10\,000~nm using logarithmic $x$ axes, while the insets show 150~nm to 300~nm using linear axes. These are shown at $\mu = 1.0$ (top) and $\mu = 0.5$ (bottom). Note the different vertical scales.}
	\label{fig:contrastspectra}
\end{figure*}

The complicated nature of the dependence of contrast on wavelength, average vertical magnetic field and limb distance can be seen in Figs.~\ref{fig:contrastvslimb} and \ref{fig:contrastspectra}. Figure~\ref{fig:contrastvslimb} shows the spatially averaged contrasts for given average vertical magnetic fields, $\langle B_z\rangle$, of 100~G (magenta), 200~G (blue), 300~G (cyan), 400~G (orange) and 500~G (red) as they vary with limb angle, for wavelengths of 214.5~nm (top), 387~nm (middle top), 601~nm (middle bottom) and 1597.5~nm (bottom). The black curves are contrasts derived from the 1D facular model used in the SATIRE solar irradiance reconstructions and is based on the FAL~P model by \citet{Fontenla1993}. As \citet{Fontenla1993} do not provide the average magnetic field strength to which the FAL~P model refers to, it is unclear to which of the simulation curves the 1D model should be compared. For that reason we have also multiplied the contrast with 0.25 (corresponding to a filling factor of 25\% of FAL~P) and have plotted these curves for comparison. The coloured shaded regions show the standard deviation of the spatially averaged contrasts calculated by propagating the standard deviations of the means of the snapshot intensities, including both the $\langle B_z \rangle$ and hydrodynamic values. The shaded regions thus provide a view of the short term temporal variation of the spatially averaged contrasts. \par
Figure~\ref{fig:contrastvslimb} shows that the mean contrast tends to increase as limb distance decreases. The steepness of these increases varies with both $\langle B_z\rangle$ and wavelength. An exception are contrasts in the UV, as illustrated in the top plot at 214.5~nm, where the mean contrast levels off at $\mu = 0.3$ for most magnetisations and begins to decrease again between $\mu = 0.3$ and $0.2$. This has been explained by the theory of magnetic flux tubes. When observed at an angle towards the limb, the hot walls of the faculae come into view and the contrast increases. The contrast reaches a maximum value and begins to decrease further towards the limb as the hot wall begins to be obscured by the granules lying immediately in front of the flux tube \citep{Spruit1976}. This effect has been observed on the Sun \citep[e.g.,][]{Auffret1990,Ortiz2002,Yeo2013}. As illustrated by the black lines, 1D atmospheric models cannot account for the obscuration and thus produce unrealistically steep contrasts near the limb. \par
Contrasts are much larger in the UV than in the visible and IR (note the difference in $y$ axis scales in Fig.~\ref{fig:contrastvslimb}). Mean contrasts at 214.5~nm reach a value of three at $\mu = 0.3$ when $\langle B_z\rangle = 500$~G, while at 601~nm in the visible the largest mean contrast remains below $0.2$. The UV also shows most variability, i.e., the range of contrasts observed across a series of snapshots is much larger than at 601 and 1597.5~nm. A more detailed view of the wavelength dependence is shown in Fig.~\ref{fig:contrastspectra}, where wavelengths of 150~nm to 10\,000~nm are displayed for disc centre, $\mu=1$ (top), and at $\mu = 0.5$ (bottom). The colour coding for the different $\langle B_z\rangle$ values is the same as in Fig.~\ref{fig:contrastvslimb}. Figure~\ref{fig:contrastspectra} illustrates the larger contrast in the UV, particularly at $\mu = 0.5$, where contrasts at $\sim 200$~nm reach levels approximately 10--20 times those seen in the visible.\par
Towards the limb, contrasts are always positive and increase for larger $\langle B_z\rangle$ values (see Figure~\ref{fig:contrastvslimb}). This holds for all wavelengths, down to approximately 170~nm, displayed in the bottom plot of Fig.~\ref{fig:contrastspectra} where the $\mu = 0.5$ contrast spectra are shown. However, towards disc centre the relationship with $\langle B_z\rangle$ is more complex (see 
Figure~\ref{fig:contrastspectra}(a)). Below 400~nm and above roughly  4000~nm, contrasts are always positive. For wavelengths between 400~nm and approximately 1500~nm, contrasts initially increase as $\langle B_z\rangle$ increases from $100$~G to $200$~G, but then decrease for stronger fields and become negative when $\langle B_z\rangle$ reaches 500~G. Contrasts for all $\langle B_z\rangle$ are very low or negative at 1.6 ${\mu}$m, close to the H$^-$ opacity minimum. The 'turnover' and sign change of the visible and NIR contrast with increasing magnetic field is not accounted for in irradiance variability models that use 1D modelled facular contrasts. In SATIRE, for example, the 1D contrasts are linearly scaled with the observed magnetic field, up to a saturated magnetic field strength. This means that larger measured magnetic fields, up to this saturated value, lead to larger (positive or negative) contrasts that always maintain their sign. This is illustrated in Fig.~\ref{fig:contrastvslimb} where the dashed and dot-dashed lines illustrate two different scalings; such linear scaling fails to capture the changing limb dependence of the contrast as a function of magnetic field.\par
In Fig.~\ref{fig:contrastspectra} some spectral features stand out. For example, a marked rise in contrast is seen between approximately 4300 and 5500~nm and, to a lesser extent, also around 2300~nm. Both regions are dominated by rovibrational CO lines \citep{Wallace1996} that are sensitive to the atmospheric structure near the temperature minimum \citep{Noyes1972,Uitenbroek1994}. A large jump is also seen at 400~nm at disc centre, which is due to the hydrogen Balmer series absorption edge (364.6~nm) and multiple absorption lines including the Ca~{\sc ii} H and K lines (396.9~nm and 393.4~nm respectively), and the CN violet system (385.8 and 388.8~nm). Below 300~nm, in the insets, multiple spectral lines are seen. Of particular note are the Mg~{\sc ii} h and k lines at 279.6~nm and 280.3~nm, that combine to form a peak at 280~nm, as well as the Mg~{\sc i} line at 285.2~nm. A large jump is seen at 208~nm at disc centre, this is an aluminium absorption edge.
\par
The contrasts derived compare well with those measured by \citet{Yeo2013}, where continuum intensity contrast with respect to the quiet Sun over a resolution element in SDO/HMI was measured near the 617.3~nm Fe spectral line. Measurements at disc centre vary between approximately -0.07 and 0.02, with larger magnetic fields generally leading to more negative contrasts. Values for our 601~nm snapshot contrasts lie between -0.02 and 0.02. Towards the limb at $\mu = 0.5$ values are found to lie in similar ranges, with larger magnetic fields having contrasts close to 0.1 for both simulations and measurements. For positions closer to the limb, the measurements indicate a decrease in contrast beyond $\mu=0.3$ that is absent in simulations at 601~nm. Differences are expected in exact values as our derived contrasts are taken by averaging over a larger area than the spatial resolution of measurements taken by \citet{Yeo2013}. Additionally, for the contrasts used here, only the initial average vertical magnetic field is considered, whereas \citet{Yeo2013} consider the line-of-sight magnetic field measured in a single spectral line, Fe I 617.3~nm, (which  corresponds to a specific atmospheric height). The magnetic fluxes in the two studies are thus not directly comparable. \par

\section{Discussion and Summary}
We have calculated LTE spectra between 149.5 and 160000~nm of magnetoconvection simulations for an early G-type star with average vertical magnetic fields of 100, 200, 300, 400 and 500~G, each for limb distances from $\mu = 0.2$ to $1.0$. By comparing these to spectra from field-free simulations, we were, for the first time, able to construct facular contrast as a function of $\mu$, $\langle B_z\rangle$ and $\lambda$ over a large wavelength range. We find that the contrast spectra display a complex dependence on magnetic field and limb distance. As the average vertical magnetic field increases, contrasts initially increase, but eventually 'turn over' and decrease, particularly at disc centre for wavelengths between 400 and 2000~nm. \par
This effect is not accounted for in currently available reconstructions of the total and spectral solar irradiance, TSI and SSI, respectively. 
Current semi-empirical solar irradiance reconstructions generally use contrasts derived from 1D facular models and need to invoke a free parameter to provide a scaling between the observable used to infer the facular surface coverage and the calculated contrasts (see \citealt{Ermolli2013}). The SATIRE model \citep{Yeo2014}, for example, scales the contrast of faculae to the observed magnetic field between the noise threshold and the saturation limit, above which the brightness of faculae saturates and does not increase further. The saturation limit is a free parameter whose value is chosen by fitting the reconstructed TSI to measured TSI values. The calculations presented in this paper provide spectral information for many pixels containing a large range of magnetised atmospheres. The contrasts presented here (see Fig.~\ref{fig:contrastspectra}) are averages over a simulation box. Many solar observatories such as the HMI instrument on SDO \citep{Schou+2012} now have greater resolution than the box size that we have examined here. The next step is thus to link our simulated spectra to the magnetic field strengths seen in typical pixels observed in magnetograms.  This will allow TSI and SSI reconstructions that do without any free parameters and hence provide a stringent test of the assumption whether solar surface magnetism really is the cause of solar irradiance variability \citep{Yeo+inprep}. \par
Stellar magnetic activity is recognised as a significant noise source for spectral observations of transiting planets. The noise is modelled by simulating exoplanet transits for stars that are covered with spots and faculae \citep[e.g.][]{Herrero2016}. Typically, the faculae are taken to emit a black-body like spectrum, resulting in facular contrasts that do not reproduce the observed wavelength and limb-angle dependence.
The facular contrasts derived in this paper can thus be used to enhance planetary transit spectroscopy by enabling the modelling of facular regions on exoplanet host stars. They can also be used to improve our understanding of micro-variability of Sun-like stars as recorded by Kepler and CoRoT, and as will be provided in future by TESS \citep{TESS2015} and PLATO \citep{PLATO2014}.\par
MURaM simulations have been performed for various spectral types from F to M, these will allow us to investigate the spectral energy distribution of magnetised regions on other stars. Having done these calculations we will be able to infer the limits of distributions of magnetic feature coverage that could account for the variability of different spectral types.

\section*{Acknowledgments}
We are grateful to Alexander Shapiro for useful discussions and helpful remarks. 
CMN would like acknowledge support through studentship funding of the UK Science and Technology Facilities Council (STFC). This work was supported by the German Federal Ministry of 
Education and Research under project 01LG1209A and partly by the BK21 plus program through the National Research Foundation (NRF) funded by the Ministry of Education of Korea. It was also supported by STFC Grants ST/N000838/1 and ST/K001051/1, and funding from the European Community's Seventh Framework Programme (FP7 2012) under grant agreement 313188 (SOLID). We would also like to thank the International Space Science Institute, Bern, for their support of science teams 335 and 373 and the resulting helpful discussions. Finally, the authors thank the referee for detailed comments that have improved the manuscript.
\titlerunning
\newpage
\newpage
\bibliographystyle{aa} 
\bibliography{PaperI.bib} 
\newpage
\appendix
\section{Additional figures}
\label{App:AddFigs}
\begin{figure*}[]
	\centering
\textit{}	\resizebox{\hsize}{!}{\includegraphics[width=\textwidth]{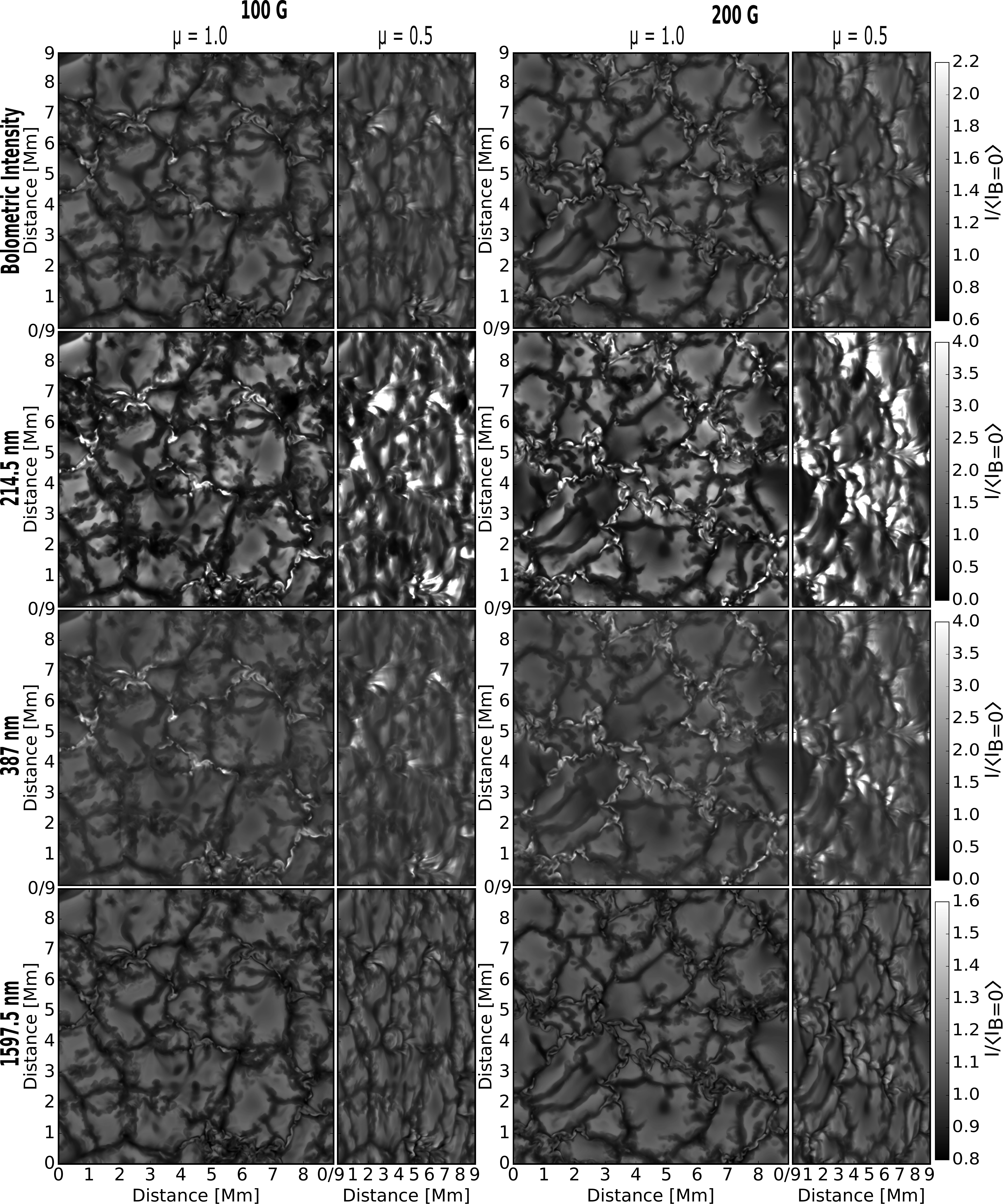}}
	\caption{Same as Fig.~\ref{fig:temp} but for $\langle B_z\rangle=100$~G and $\langle B_z\rangle=200$~G.}
		\label{fig:spectralimages100200}
\end{figure*}

\begin{figure*}[]
	\centering
\textit{}	\resizebox{\hsize}{!}{\includegraphics[width=\textwidth]{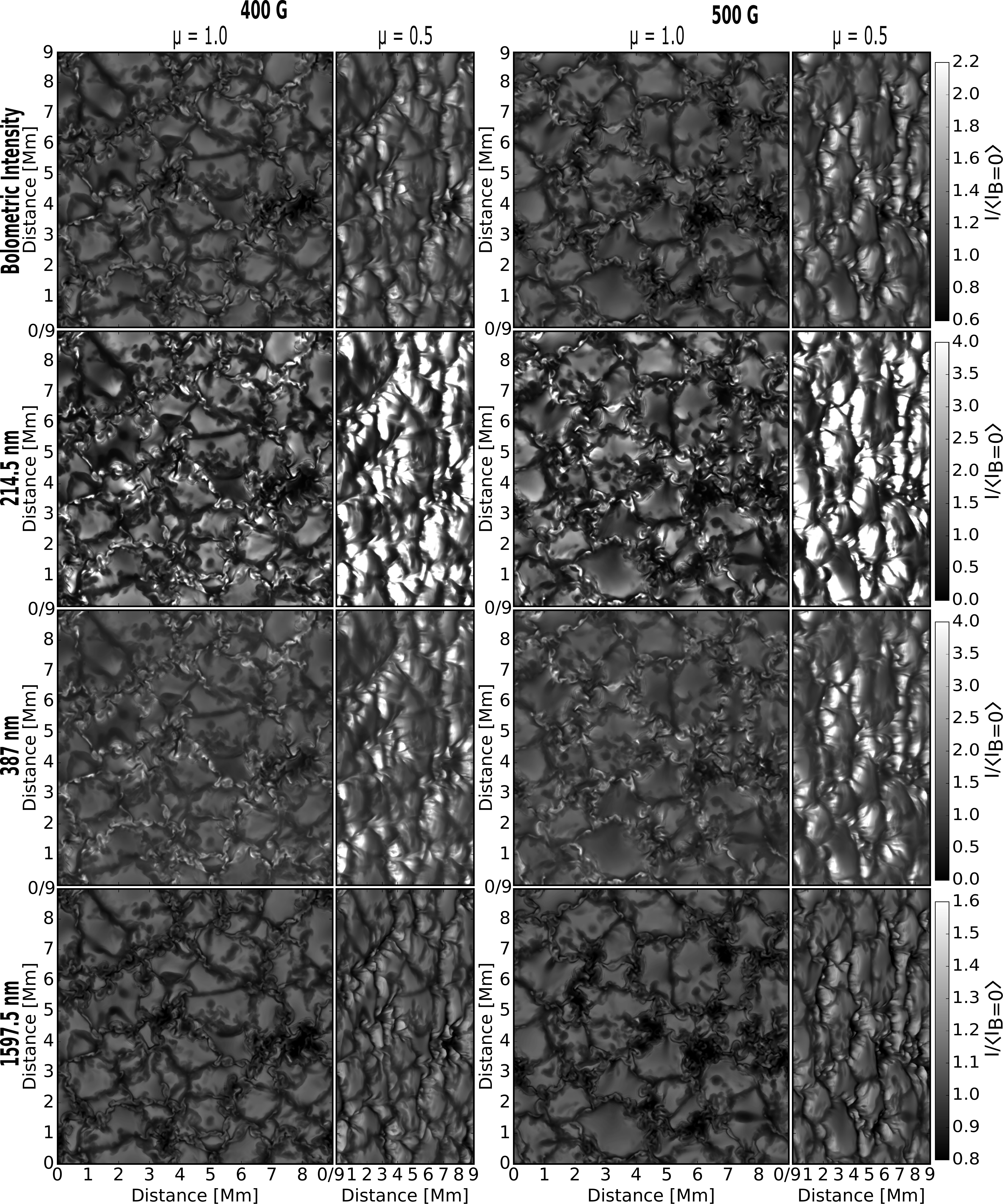}}
	\caption{Same as Fig.~\ref{fig:temp} but for $\langle B_z\rangle=400$~G and $\langle B_z\rangle=500$~G.}
		\label{fig:spectralimages400500}
\end{figure*}

Figures \ref{fig:spectralimages100200} and \ref{fig:spectralimages400500} show snapshots from initial average magnetic field simulations of 100~G and 200~G as well as 400~G and 500~G, respectively. We show bolometric images (top row) along with images for wavelength bins centred at 214.5~nm, 387~nm and 1597.5~nm. Figures~\ref{fig:spectralimages100200} and \ref{fig:spectralimages400500} complement Fig.~\ref{fig:spectralimages} in the main text.
\end{document}